\documentclass[journal]{IEEEtran}

\usepackage{graphicx}      % include this line if your document contains figures
\def\be{\begin{equation}}
\def\ee{\end{equation}}
\def\ba{\begin{array}}
\def\ea{\end{array}}
\usepackage[gen]{eurosym}
\usepackage{epstopdf}
\usepackage{dsfont}

\usepackage{multicol}
\usepackage{pgf}
\usepackage{tikz}
\usetikzlibrary{arrows,automata}
\usepackage{amsmath} % assumes amsmath package installed
\usetikzlibrary{arrows,shadows,calc,decorations.shapes,decorations} % for pgf-umlsd
\usepackage[underline=true,rounded corners=false]{pgf-umlsd}
\usepackage[american,cute inductors,smartlabels]{circuitikz}
\usepackage{mathrsfs}
\usepackage{subfigure}
\usepackage{blox}
\usepackage{graphicx,color}
\usepackage{latexsym}
\usepackage{booktabs}
\usepackage{amsmath} % assumes amsmath package installed
\usepackage{amssymb}  % assumes amsmath package installed
\usepackage{amsmath,amssymb,amsfonts}
\DeclareMathOperator{\sgn}{sgn}

\DeclareMathOperator{\diag}{diag}
\usepackage{mathtools}	
\DeclarePairedDelimiter{\abs}{\lvert}{\rvert}

\newtheorem{remark}{Remark}
\newtheorem{definition}{Definition}

\newtheorem{theorem}{Theorem}
\newtheorem{objective}{Objective}
\newtheorem{lemma}{Lemma}

\newtheorem{assumption}{Assumption}

\usepackage{siunitx}	
\usepackage{balance}	
\usepackage{tikz}

\DeclareMathAlphabet{\mathpzc}{OT1}{pzc}{m}{it}
\DeclareMathAlphabet{\mathcal}{OMS}{cmsy}{m}{n}
\newtheorem{innercustomgeneric}{\customgenericname}
\providecommand{\customgenericname}{}
\newcommand{\newcustomtheorem}[2]{%
  \newenvironment{#1}[1]
  {%
   \renewcommand\customgenericname{#2}%
   \renewcommand\theinnercustomgeneric{##1}%
   \innercustomgeneric
  }
  {\endinnercustomgeneric}
}

\newcustomtheorem{customobj}{Objective}

\newcommand{\R}{\mathds{R}}

\usepackage{scalerel}
\def\sq{\diamond}
\title{
Passivity based design of sliding modes for optimal Load Frequency Control$^{\star}$}
\author{Sebastian Trip$^{1}$, Michele Cucuzzella$^{2}$, Claudio De Persis$^{1}$, Arjan van der Schaft$^{3}$, Antonella Ferrara$^{2}$% <-this % stops a space
\thanks{$\star${This work is part of the research programme ENBARK+ with project number 408.urs+.16.005, which is (partly) financed by the Netherlands Organisation for Scientific Research (NWO). Also, this work is part of the EU Project `ITEAM' (project reference: 675999). Preliminary results have appeared in \cite{cucuzzella_2017_acc}.
}
}% <-this % stops a space%
\thanks{$^{1}$Sebastian Trip and Claudio De Persis are with ENTEG, Faculty of Science and Engineering, University of Groningen, Nijenborgh 4, 9747 AG Groningen, the Netherlands. {\tt\small \{s.trip, c.de.persis\}@rug.nl}.} %} \\
\thanks{$^{2}$Michele Cucuzzella and Antonella Ferrara are with the Dipartimento di Ingegneria Industriale e dell'Informazione, University of Pavia, via Ferrata 1, 27100 Pavia, Italy. {\tt\small michele.cucuzzella@gmail.com, antonella.ferrara@unipv.it}.}
\thanks{$^{3}$Arjan van der Schaft is
 with the Johann Bernoulli Institute for Mathematics
and Computer Science, Faculty of Science and Engineering, University of Groningen, Nijenborgh 9, 9747 AG
Groningen, the Netherlands. {\tt\small a.j.van.der.schaft@rug.nl}.}
}

\begin{document}
\maketitle
%\thispagestyle{empty}
%\pagestyle{empty}

%\begin{document}
%\begin{frontmatter}
%
%\title{An energy function based design \\ of second order sliding modes  \\ for Automatic Generation Control\thanksref{footnoteinfo}}
%% Title, preferably not more than 10 words.
%
%\thanks[footnoteinfo]{This work is partially supported by the Danish Council for Strategic Research (contract no. 11-116843) within the `Programme Sustainable Energy and Environment', under the `EDGE' (Efficient Distribution of Green Energy) research project and by the EU Project `ITEAM' (project reference: 675999).}
%
%\author[First]{Sebastian Trip}
%\author[Second]{Michele Cucuzzella}
%\author[Second]{Antonella Ferrara}
%\author[First]{Claudio De Persis}
%
%\address[First]{ENTEG, Faculty of Mathematics and Natural Sciences, University of Groningen, Nijenborgh 4, 9747 AG Groningen, the Netherlands. (e-mail: s.trip@rug.nl; c.de.persis@rug.nl).}
%\address[Second]{Dipartimento di Ingegneria Industriale e dell'Informazione, University of Pavia, via Ferrata 1, 27100 Pavia, Italy. \\(e-mail: michele.cucuzzella@gmail.com; a.ferrara@unipv.it)}
%\address[Third]{Electrical Engineering Department,
%   Seoul National University, Seoul, Korea, (e-mail: author@snu.ac.kr)}

\begin{abstract}                % Abstract of not more than 250 words.
This paper proposes a distributed sliding mode control strategy for \emph{optimal} Load Frequency Control (OLFC) in power networks, where besides frequency regulation also minimization of generation costs is achieved (economic dispatch).
%Because of unknown loads and possible network parameters uncertainties, the sliding mode control methodology is particularly appropriate for the considered control problem.
 We study a nonlinear power network partitioned into control areas, where each area is modelled by an equivalent generator including voltage and second order turbine-governor dynamics. The turbine-governor dynamics suggest the design of a sliding manifold, such that the turbine-governor system enjoys a suitable passivity property, once the sliding manifold is attained. This work offers a new perspective on OLFC by means of sliding mode control, and in comparison with existing literature, we relax required dissipation conditions on the generation side and assumptions on the system parameters.
\end{abstract}


\vspace{-0.3cm}
\section{Introduction}
%\todo{Work of Sira ramirez}
%\todo{Add work of Orlov and Utkin}
A power mismatch between generation and demand gives rise to a frequency in the power network that can deviate from its nominal value. Regulating the frequency back to its nominal value by Load Frequency Control (LFC) is challenging and it is uncertain if current implementations are adequate to deal with an increasing share of renewable energy sources~\cite{apostolopoulou_2016_tps}.
%\smallskip \\
Traditionally, the LFC is performed at each control area by a primary droop control and a secondary proportional-integral (PI) control.
To cope with the increasing uncertainties affecting a control area and to improve the controller's performance, advanced control techniques have been proposed to redesign the conventional LFC schemes, such as model predictive control (MPC) \cite{7065326}, adaptive control \cite{Zribi2005575}, fuzzy control~\cite{Chang1997145} and sliding mode (SM) control. However, due to the predefined power flows through the tie-lines, the possibility of achieving economically optimal LFC is lost \cite{4077135}.
Besides improving the stability and the dynamic performance of power systems, new control strategies are additionally required to reduce the operational costs of LFC \cite{lai2001power}. In this paper we propose a novel distributed \emph{optimal} LFC (OLFC) scheme that incorporates the economic dispatch into the LFC loop, departing from the conventional tie-line requirements. An up-to-date survey on recent results on offline and online optimal power flows and OLFC can be found in \cite{molzahn_2017_tsg}. We restrict ourselves here to a brief overview of online solutions to OLFC that are close to the presented work. Particularly, we focus on distributed solutions, in contrast to more centralized control schemes that have been studied in e.g. \cite{trip_2017_ifac,DORFLER2017296,xi2017power}.
%\smallskip \\
In order to obtain OLFC, the vast majority of distributed solutions appearing in the literature fit in one of two categories. First, the economic dispatch problem is distributively solved by a primal-dual algorithm converging to the solution of the associated Lagrangian dual problem
\cite{zhang_2015_automatica, chen2015, stegink_2016_arxiv, you_2014_cdc, kasis_2016_arxiv, jokic_2009_epes, mudumbai_2012_ps, miao_2016_tps, cai_2015_cdc, apostolopoulou_2015_naps,Yi201545,Yi2016259}.
%\cite{zhang_2015_automatica, chen_2015_cns, stegink_2016_arxiv, you_2014_cdc,  miao_2016_tps, cai_2015_cdc, apostolopoulou_2015_naps}.
This approach generally requires measurements of the loads or the power flows, which is not always desirable in a LFC scheme. This issue is avoided by the second class of solutions, where a distributed consensus algorithm is employed to converge to a state of identical marginal costs, solving the  economic dispatch problem in the unconstrained case
\cite{burger.et.al.mtns14b,trip_2016_automatica,schiffer_2016_ecc,zhao_2015_acc,xi2017multi,monshizadeh_2015b_arxiv,andreasson_2013_ecc,kar_2012_pesgm,binetti_2014_tps,rahbari_2014_tsg,yang_2013_tps, yang2016, zhang_2012}.
%\cite{burger.et.al.mtns14b,trip_2016_automatica,zhao_2015_acc,kar_2012_pesgm,binetti_2014_tps,rahbari_2014_tsg,yang_2013_tps}.
 The proposed sliding mode controller design in this work is compatible with both approaches, although we put the emphasize on a distributed consensus based solution and remark on the primal-dual based approach.
\subsection{Main contributions}
%\todo{Add general comment about passivity and sliding modes}
 Sliding mode control
  %\cite{Utkin,CESKS}
  %\todo{Can we replace one of the references below?. We have twice the same authors.}
  has been used to improve the conventional LFC schemes \cite{Vrdoljak2010514}, possibly together with
%fuzzy logic \cite{725839} and
disturbance observers \cite{Mi2016446}. However, the proposed use of SM to obtain a distributed OLFC scheme is new and can offer a few advantages over the previous results on OLFC. Foremost, it is possible to incorporate the widely used second order model for the turbine-governor dynamics that is generally neglected in the analytical OLFC studies. Since the generated control signals in OLFC schemes adjust continuously and in real-time the governor set points, it is important to incorporate the generation side in a satisfactory level of detail. In this paper, we adopt a \emph{nonlinear} model of a power network, including voltage dynamics, partitioned into control areas having an arbitrarily complex and meshed topology. The generation side is modelled by an equivalent generator including voltage dynamics and second order turbine-governor dynamics, which is standard in studies on conventional LFC schemes. We propose a \emph{distributed} SM controller that is shown to achieve frequency control, while minimizing generation costs.
%This result is obtained by avoiding the measurement of the power demand and the use of observers, which is an element concurring to the ease of practical implementation of the proposed control strategy.
The proposed control scheme continuously adjusts the governor set point. Conventional SM controllers can suffer from the notorious drawback known as chattering effect, due to the discontinuous control input. To alleviate this issue, we incorporate the well known Suboptimal Second Order Sliding Mode (SSOSM) control algorithm \cite{661074} leading to a continuous control input.
%\smallskip \\
 To design the controllers obtaining OLFC, we recall an incremental passivity property of the power network \cite{trip_2016_automatica} that prescribes a suitable sliding manifold. Particularly, the non-passive turbine-governor system, constrained to this manifold, is shown to be incrementally passive allowing for a passive feedback interconnection, once the closed-loop system evolves on the sliding manifold. The proposed approach differs substantially from two notable exceptions that also incorporate the turbine-governor dynamics (\cite{trip_2017_tns}, \cite{kasis2017stability}) and shows some benefits. In contrast to \cite{trip_2017_tns}, we do not impose constraints on the permitted system parameters, and in contrast to \cite{kasis2017stability} we do not impose dissipation assumptions on the generation side and allow for a higher relative degree (see also Remark \ref{rem:contri}). Furthermore, we believe that the chosen approach, where the design of the sliding manifold is inspired by desired passivity properties, offers new perspectives on the control of networks that have similar control objectives as the one presented, e.g. achieving power sharing in microgrids. As this paper is (to the best of our knowledge) the first to use sliding mode control to obtain OLFC, it  additionally enables further studies to compare the performance with respect to other approaches found in the literature.
%\smallskip \\
\subsection{Outline}
The present paper is organized as follows. In Section \ref{sec:model} the network model is introduced. In Section \ref{sec3} the considered OLFC problem is formulated. The proposed controller is described and analyzed in Section \ref{sec4} and \ref{sec5}, respectively. Simulation results are reported and discussed in Section \ref{sec6}, while some conclusions are finally gathered in Section~\ref{sec7}.
%
%
%
%
%%%%%%%%%% PROBLEM FORMULATION
\section{Nonlinear power network model}
\label{sec:model}
\begin{figure}[t]
        \centering
        \begin{tikzpicture}[scale=0.8, transform shape]

        \begin{small}
        \bXInput{input}

        \begin{tiny}
        \bXCompSum[4.6]{sum1}{input}{-}{}{+}{}
        \end{tiny}

        \bXLink[$u_i$]{input}{sum1}

        \bXStyleBloc{rounded corners,fill=blue!10,text=black}
        \bXBloc[2.5]{governor}{$\frac{1}{T_{g_i}\,s\;+\;1}$}{sum1}
        \begin{scriptsize}
        \bXLinkName[2.8]{governor}{Governor$_i$}
        \end{scriptsize}

        \bXLink[]{sum1}{governor}

        \bXStyleBloc{rounded corners}
        \bXBloc[2.5]{turbine}{$\frac{1}{T_{t_i}\,s\;+\;1}$}{governor}
        \begin{scriptsize}
        \bXLinkName[2.8]{turbine}{Turbine$_i$}
         \end{scriptsize}

        \bXLink[$  P_{g_i}$]{governor}{turbine}

        \begin{tiny}
        \bXCompSum[8.4]{sum2}{turbine}{-}{-}{+}{}
        \end{tiny}

        \bXBranchy[-4]{sum2}{d}
        \bXLink{d}{sum2}
        \bXLinkName[0.5]{d}{$  P_{d_i}$}
         \bXBranchy[4]{sum2}{d12}
        \bXLink{d12}{sum2}
        \bXLinkName[-0.5]{d12}{$B_{ij} V_{i} \,  V_{j} \sin{(  \delta_i  -   \delta_j )}$}

        \bXLink[$  P_{t_i}$]{turbine}{sum2}

        \bXStyleBloc{rounded corners}
        \bXBloc[1.8]{power_system}{$\frac{K_{p_i}}{T_{p_i}\,s\;+\;1}$}{sum2}
        \begin{scriptsize}
        \bXLinkName[2.8]{power_system}{Power System$_i$}
        \end{scriptsize}

        \bXLink[]{sum2}{power_system}

        \bXOutput[3]{output}{power_system}

        \bXLink[]{power_system}{output}
        \bXLinkName[0.8]{output}{$  f_i$}

        \bXBranchy[-7]{power_system-output}{reg_primaria}

         \bXStyleBloc{rounded corners,fill=black!10,text=black}

        \bXChainReturn[16.25]{reg_primaria}{r1/$\frac{1}{R_i}$}

        \bXLinkyx{power_system-output}{r1}
        \bXLinkxy{r1}{sum1}

       \bXBranchy[+9.0]{input}{input_j}

        \begin{tiny}
        \bXCompSum[4.6]{sum1_j}{input_j}{}{-}{+}{}
        \end{tiny}

        \bXLink[$u_j$]{input_j}{sum1_j}

        \bXStyleBloc{rounded corners,fill=blue!10,text=black}
        \bXBloc[2.4]{governor_j}{$\frac{1}{T_{g_j}\,s\;+\;1}$}{sum1_j}
        \begin{scriptsize}
        \bXLinkName[-2.8]{governor_j}{Governor$_j$}
        \end{scriptsize}

        \bXLink[]{sum1_j}{governor_j}

        \bXStyleBloc{rounded corners}
        \bXBloc[2.45]{turbine_j}{$\frac{1}{T_{t_j}\,s\;+\;1}$}{governor_j}
        \begin{scriptsize}
        \bXLinkName[-2.8]{turbine_j}{Turbine$_j$}
        \end{scriptsize}

        \bXLink[$  P_{g_j}$]{governor_j}{turbine_j}

        \begin{tiny}
        \bXCompSum[8.4]{sum2_j}{turbine_j}{+}{-}{+}{}
        \end{tiny}

        \bXBranchy[4]{sum2_j}{d_j}
        \bXLink{d_j}{sum2_j}
        \bXLinkName[-0.5]{d_j}{$  P_{d_j}$}
          \bXBranchy[-4]{sum2_j}{d21}
        \bXLink{d21}{sum2_j}

        \bXLink[$  P_{t_j}$]{turbine_j}{sum2_j}

        \bXStyleBloc{rounded corners}
        \bXBloc[1.75]{power_system_j}{$\frac{K_{p_j}}{T_{p_j}\,s\;+\;1}$}{sum2_j}
        \begin{scriptsize}
        \bXLinkName[-2.8]{power_system_j}{Power System$_j$}
        \end{scriptsize}

        \bXLink[]{sum2_j}{power_system_j}

        \bXBranchx[5.13]{power_system_j}{output2}

        \bXLink[]{power_system_j}{output2}
        \bXLinkName[0.8]{output2}{$  f_j$}

       \bXBranchy[7]{power_system_j-output2}{reg_primaria_j}

         \bXStyleBloc{rounded corners,fill=black!10,text=black}

        \bXChainReturn[16.25]{reg_primaria_j}{r1j/$\frac{1}{R_j}$}

        \bXLinkyx{power_system_j-output2}{r1j}
        \bXLinkxy{r1j}{sum1_j}
	\end{small}

        \end{tikzpicture}
        \caption{Block diagram of two interconnected control areas. The voltage dynamics are omitted. }
        \label{fig:2areas}
\end{figure}
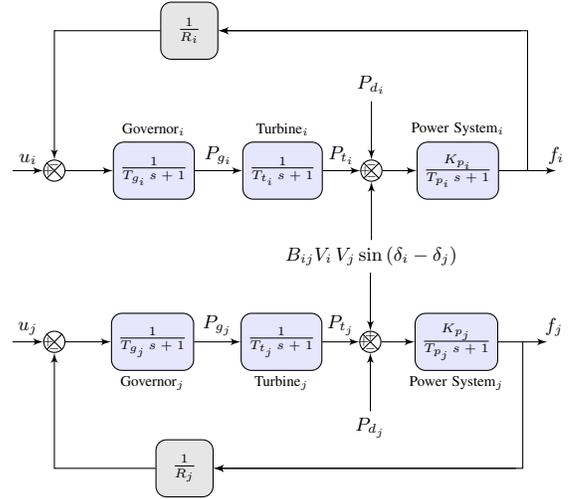
%In this section the dynamic model of a power network partitioned into control areas is presented. The dynamic behaviour of a single control area is described by an equivalent thermal power plant with a non-reheat turbine, which is commonly represented by second order turbine-governor dynamics.

Consider a power network consisting of $n$ interconnected control areas. The network topology is represented by a connected and undirected graph $\mathcal{G} = (\mathcal{V},\mathcal{E})$, where the nodes $\mathcal{V} = \{1,...,n\}$, represent the control areas and the edges $\mathcal{E}  = \{1,...,m\}$, represent the transmission lines connecting the areas. The topology can be described by its corresponding incidence matrix $\mathcal{B} \in \R^{n \times m}$. Then, by arbitrarily labeling the ends of edge $k$ with a $+$ and a $-$, one has that
\begin{equation*}
\label{eq:incidence}
\mathcal{B}_{ik}=
\begin{cases}
+1 \quad &\text{if $i$ is the positive end of $k$}\\
-1 \quad &\text{if $i$ is the negative end of $k$}\\
~~0 \quad &\text{otherwise}.
\end{cases}
\end{equation*}
A control area is represented by an equivalent generator and a load, where the governing dynamics of the $i$-th area are described by the so called `flux-decay' or `single-axis model' given as\footnote{~For notational simplicity, the dependency of the variables on time $t$ is omitted throughout most of this paper.} \cite{machowski_power_2008}:
\begin{align}
\dot{\delta}_i  = &~ f_{i}\nonumber \\
T_{pi}\dot{f}_i  = &- f_i  +  K_{pi} \Big( P_{ti} - P_{di} + \sum_{j \in \mathcal{N}_i}^{} V_{i} V_{j} B_{ij}\sin{(  \delta_i  -   \delta_j )} \Big) \nonumber \\
T_{Vi} \dot V_i =&~ \overline E_{fi} - \big(1- (X_{di} - X_{di}')B_{ii}\big)V_i \label{eq:plant_i1} \\
&-(X_{di} - X_{di}')\sum_{j \in \mathcal{N}_i}^{} V_{j} B_{ij}\cos{(  \delta_i  -   \delta_j )}, \nonumber
\end{align}
%where $P_i$ and $Q_i$ are the active and reactive power given by
%\begin{align}\label{power}
  %\begin{split}
   % P_i =&   \displaystyle  \\
%    Q_i =&~ \frac{V_i^2}{X_{ii}} + \sum_{j \in \mathcal{N}_i}\frac{ V_{i} V_{j}} {X_{ij}} \cos{(  \delta_i  -   \delta_j )},
%  \end{split}
%\end{align}
where $\mathcal{N}_i$  is the set of control areas connected to the $i$-th area by transmission lines.
Note that we assume that the network is lossless, which is generally valid in high voltage transmission networks where the line resistance is negligible.
Moreover, $P_{ti}$ in \eqref{eq:plant_i1} is the power generated by the $i$-th (equivalent) plant and can be expressed as the output of the following second order dynamical system that describes the behaviour of both the governor and the turbine:
\begin{align}
\begin{split}
\label{eq:plant_i2}
T_{ti} \dot{P}_{ti}  = &-P_{ti}  +  P_{gi} \\
T_{gi} \dot{P}_{gi}  = &-\frac{1}{R_i}	  f_i  - P_{gi}   + u_i.
\end{split}
\end{align}
The symbols used in  \eqref{eq:plant_i1} and \eqref{eq:plant_i2} are described in Table~\ref{tab:symbols}. To further illustrate the dynamics, a block diagram for a two area network is provided in Figure 1. In this paper we aim at the design of a continuous control input $u_i$ to achieve both frequency regulation and economic efficiency (optimal Load Frequency Control).
\begin{table}
\centering
\caption{Description of the used symbols}
{\begin{tabular}{ll}
\toprule			
											&State variables\\					
\midrule
		$\delta_i$										& Voltage angle\\
		$f_i$											& Frequency deviation\\
$ V_i$ & Voltage \\
		$P_{ti}$								& Turbine output power \\
		$P_{gi}$								& Governor output\\
\midrule
& Parameters\\
\midrule
		$T_{pi}$								& Time constant of the control area\\
		$T_{ti}$								& Time constant of the turbine\\
		$T_{gi}$								& Time constant of the governor\\
		$T_{Vi}$								& Direct axis transient open-circuit constant\\
		$K_{pi}$								& Gain of the control area\\
		$R_i$										& Speed regulation coefficient\\
$X_{di}$ & Direct synchronous reactance \\
$X'_{di}$ & Direct synchronous transient reactance \\
		$B_{ij}$										&Transmission line susceptance\\
\midrule
& Inputs \\
\midrule
		$u_i$										& Control input to the governor\\
		$\overline E_{fi}$									& Constant exciter voltage\\
		$P_{di}$								    & Unknown power demand\\
\bottomrule
\end{tabular}}
\label{tab:symbols}
\end{table}
%\smallskip \\
To study the power network we write system \eqref{eq:plant_i1} compactly for all areas $i \in \mathcal{V}$ as
%\todo{check the sign in the second line}
\begin{align}\label{syscompact}
\begin{split}
\dot{\eta}  = & ~\mathcal{B}^{T}f  \\
T_p\dot{f}  = &- f  +  K_p(P_{t}  - P_{d} {\color{black}{-}} \mathcal{B}  \Gamma(V) \sin(  \eta ) )\\
T_V \dot V =& -(X_d - X'_d)E(\eta)V + \overline E_{f}, \\
\end{split}
\end{align}
and the
 turbine-governor dynamics in \eqref{eq:plant_i2} as
\begin{align}
\begin{split}\label{tgcompact}
T_t\dot{P}_{t}  = &-P_{t}  +  P_{g} \hspace{3.8cm}\\
T_g\dot{P}_{g}  = &-R^{-1} f  - P_{g}   +  u,
\end{split}
\end{align}
 where $  \eta = \mathcal{B}^{T}   \delta \in \R^m$ is vector describing the differences in voltage angles. Furthermore, $\Gamma = \diag\{ \Gamma_1, \dots, \Gamma_m \}$, where $\Gamma(V)_k =  V_{i} V_{j}B_{ij}$, with $k \sim \{i,j \}$, i.e., line $k$ connects areas $i$ and $j$. The components of the matrix $E(\eta) \in \R^{n \times n}$ are defined as
 \begin{align}
   \begin{split}
 E_{ii}(\eta) =& \frac{1}{X_{di} - X^{'}_{di}} - B_{ii} \hspace{3.2em} i \in \mathcal{V} \\
     E_{ij}(\eta) =&~ B_{ij}\cos(\eta_k) = E_{ji}(\eta) \hspace{1.2em} k \sim \{i,j\} \in \mathcal{E} \\
   E_{ij}(\eta) =&~0 \hspace{9.2em}~ \text{otherwise}.
   \end{split}
 \end{align}
 The remaining symbols follow straightforwardly from \eqref{eq:plant_i1} and (\ref{eq:plant_i2}), and are vectors and matrices of suitable dimensions.
  %Since for realistic networks $X_{di}>X_{di^{'}}$ and $B_ii $
%where $  \eta = \mathcal{B}^{T}   \delta \in \R^m$, $  f \in \R^n$, $  P_t \in \R^n$, $  P_g \in \R^n$, $\Gamma = \diag\{ \Gamma_1, \dots, \Gamma_m \}$, with $\Gamma(V)_k =  \frac{V_{i} V_{j}}{ X_{ij}}$, where line $k$ connects areas $i$ and $j$, $\sin(  \eta) = (\sin(  \eta_1), \dots, \sin(  \eta_m))^T$, $  P_d \in \R^n$ and $  u \in \R^n$. Matrices $T_p, T_t, T_g, K_p, R$ are suitable $n \times n$ diagonal matrices, e.g., $K_p = \diag \{ K_{p1}, \dots, K_{pn}  \}$.
\bigskip
\begin{remark}({\bf{Reactance and susceptance}})
  For each (equivalent) generator $i \in \mathcal{V}$, the reactance is higher than the transient reactance, i.e. $X_{di} > X'_{di}$ \cite{kundur1994power}. Furthermore, the self-susceptance of area $i \in \mathcal{V}$ is given by $B_{ii} = \sum_{j \in \mathcal{N}_i} B_{ij}$ and the susceptance of a line satisfies $B_{ij} = B_{ji} < 0$. Consequently, $E(\eta)$ is a strictly diagonally dominant and symmetric matrix with positive elements on its diagonal and is therefore positive definite. $\hfill \sq$
\end{remark}
\bigskip
To permit the controller design in the next sections, the following assumption is made on the \emph{unknown} demand (unmatched disturbance) and the available measurements:
%\smallskip
\bigskip
\begin{assumption}
\label{ass:0}({\bf{Available information}})
The variables $f_i,P_{ti}$ and $P_{gi}$ are locally available at control area $i$. The unmatched disturbance $ P_{di}$ is unknown, and can be bounded as
$
\abs{P_{di}} \leq \mathcal{D}_i,
$
where $\mathcal{D}_i$ is a positive constant available at control area $i$. $\hfill \sq$
\end{assumption}
\bigskip
{\color{black}{In case not all variables are locally available, Assumption \ref{ass:0} can be relaxed by implementing observers that estimate the unmeasured states in a finite time (see for instance \cite{Rinaldi2017}).}}
\section{Incremental passivity of the power network}\label{sec3}
In this section we recall a useful incremental passivity property of system (\ref{syscompact}) that has been established before in \cite{trip_2016_automatica}.
To facilitate the discussion, we first define `incremental passivity'.
%\smallskip
\bigskip
\begin{definition}({\bf Incremental passivity})
  System \begin{align} \begin{split}
  \dot x =&~ \zeta(x,u)\\
   y=&~h(x),
  \end{split}
  \end{align}
  $x \in \mathds{R}^n$, $u,y \in \mathds{R}^m$, is incrementally passive with respect to\footnote{~We state the incremental passivity property with respect to a steady state solution, and not with respect to any solution.} a constant triplet $(\overline x, \overline u, \overline y)$ satisfying
  \begin{align}
  \begin{split}
    \boldsymbol{0} =& ~\zeta(\overline x, \overline u) \\
    \overline y =& ~h(\overline x),
    \end{split}
  \end{align}
  if there exists a continuously differentiable function $\mathcal{S} : \mathds{R}^n \rightarrow \mathds{R}_+$, such that for all $x \in \mathds{R}^n$, $u \in \mathds{R}^m$ and $y=h(x)$, $\overline y = h(\overline x)$
\begin{align}
\begin{split}
  \dot{\mathcal{S}} = \frac{\partial \mathcal{S}}{\partial x} \zeta(x,u) + \frac{\partial \mathcal{S}}{\partial \overline x} \zeta (\overline x,\overline u) \leq &-W(y, \overline y) \\ &+ (y-\overline y)^T (u-\overline u).
\end{split}
\end{align}
In case $W(y,\overline y) > 0$,  the system is called `\emph{output strictly} incrementally passive'. In case $\mathcal{S}$ is not lower bounded, the system is called `incrementally \emph{cyclo-}passive'. $\hfill \sq$
\end{definition}
\bigskip
Before we can establish this incremental passivity property for the considered power network model, we first need the following assumption on the existence of a steady state solution.
%\smallskip
\bigskip
\begin{assumption}\label{a2}({\bf Steady state solution})
The unknown power demand (unmatched disturbance) $P_{d}$ is constant and for a given $\overline P_t$, there exist a $\overline u$ and state $( \overline \eta, \overline f, \overline V, \overline P_t, \overline P_g)$  that satisfies
  \begin{align}%\label{assum_ss}
    \begin{split}
\mathbf{0}  = & ~\mathcal{B}^{T}\overline f \label{sseq} \\
\mathbf{0} = &- \overline f + K_p(\overline   P_{t}  -  P_{d} - \mathcal{B}  \Gamma(\overline V) \sin(  \overline \eta )) \\
\mathbf{0} =& -(X_d - X'_d)E(\overline \eta)\overline V + \overline E_f,
\end{split}
\end{align}
and
\begin{align}
\begin{split}\label{sstg}
\mathbf{0} = &-\overline P_{t}  +  \overline P_{g} \hspace{10.5em}\\
\mathbf{0}= &-R^{-1} \overline f - \overline P_{g}   +  \overline u.
    \end{split}
  \end{align}
  $ \hfill \sq$
\end{assumption}
\bigskip
%\smallskip
To state an incremental passivity property of (\ref{syscompact}), we make use of the following storage function \cite{trip_2016_automatica}, \cite{persis_2016_arxiv_microgrid}:
\begin{align}
\begin{split} \label{sf}
  S_1(\eta, f, V) =&~ \frac{1}{2}f^T T_p f + \frac{1}{2}V^T E(\eta)V,
  \end{split}
\end{align}
that can also be interpreted as a Hamiltonian function of the system \cite{stegink_2016_arxiv}.
%\smallskip
\bigskip
\begin{lemma}({\bf Incremental cyclo-passivity of (\ref{syscompact})})
  System (\ref{syscompact}) with input $P_t$ and output $f$ is an  output strictly incrementally cyclo-passive system, with respect to the constant $(\overline \eta, \overline f, \overline V)$ satisfying (\ref{sseq}).
\end{lemma}
 \vspace{0.2cm}
\begin{IEEEproof}
For notational convenience we define $x = (\eta, f,  V)$.
  A tedious but straightforward evaluation of (note the use of a calligraphic $\mathcal{S}$)
  \begin{align}\label{incrementalsf}
  \begin{split}
   \mathcal{S}_1(x) =  S_1(x) - S_1(\overline x)  - \nabla S_1(\overline x)^T(x - \overline x),
   \end{split}
  \end{align}
  shows that $\mathcal{S}_1(x)$ satisfies \cite{trip_2016_automatica}, \cite{persis_2016_arxiv_microgrid}
  \begin{align}
  \begin{split}
    \dot{ \mathcal{S}}_1 (x) =& -f^TK_p^{-1}f  - \dot V^T T_V(X_d - X'_d)^{-1} \dot V \\ &+ (f-\overline f)^T(P_t - \overline P_t),
     \end{split}
  \end{align}
  along the solutions to (\ref{syscompact}).
\end{IEEEproof}
%\smallskip
\bigskip
For the stability analysis in Section \ref{sec6} the following technical assumption is needed on the steady state that eventually allows us to infer boundedness of solutions.\footnote{~In case boundedness of solutions can be inferred by other means, Assumption \ref{assum:voltage} can be omitted.}
%\smallskip
%\todo{Check, I copied this from Tjerks paper, but I don't understand some terms.}
\bigskip
\begin{assumption}\label{assum:voltage}({\bf Steady state voltages and voltage angles})
Let $\overline V \in \mathds{R}^n_{>0}$ and let differences in steady state voltage angles satisfy
\begin{align}
\overline \eta_k \in (-\frac{\pi}{2}, \frac{\pi}{2}) \quad \forall k \in \mathcal{E}.
\end{align}
Furthermore, for all $i \in \mathcal{V}$ it holds that
%\todo{check signs}
\begin{align}
\begin{split}
  &\frac{1}{X_{di} - X'_{di}} - B_{ii} + \sum_{k \sim \{i,j \} \in \mathcal{E}} \frac{B_{ij}(\overline{V}_i + \overline{V}_j \sin^2(\overline \eta_k))}{\overline{V}_i\cos(\overline \eta_k)} > 0.\\
  %&>\sum_{k\sim \{i,j\} \in \mathcal{E}} \frac{V_j \cos(\overline \eta_k)}{X_{ij}}\Big(1 + \frac{ \overline V_i}{\overline V_j} \tan^2(\overline \eta_k) \Big) > 0.
  \end{split}
\end{align}
$\hfill  \sq$
\end{assumption}
%\smallskip
\bigskip
The assumption above holds if the generator reactances are small compared to the line reactances and the differences in voltage (angles) are small \cite{persis_2016_arxiv_microgrid}. It is important to note that this holds for typical operation points of the power network.
The main consequence of Assumption \ref{assum:voltage} is that the incremental storage function $\mathcal{S}_1$ now obtains a strict local minimum at a steady state satisfying (\ref{sseq}).
%\smallskip
%\todo{Add remark of positive voltages}
\bigskip
\begin{lemma}({\bf Local minimum of $\mathcal{S}_1$})
  Let Assumption 3 hold. Then, the incremental storage function $\mathcal{S}_1$ has a local minimum at $(\overline \eta, \overline f, \overline V)$ satisfying (\ref{sseq}).
  \end{lemma}
    \vspace{0.2cm}
  \begin{IEEEproof}
     Under Assumption \ref{assum:voltage}, the Hessian of (\ref{sf}), evaluated at $(\overline \eta, \overline f, \overline V)$, is positive definite \cite[Lemma 2]{trip_2016_automatica}, \cite[Proposition 1]{persis_2016_arxiv_microgrid}. Consequently, $S_1$ is strictly convex around $(\overline \eta, \overline f, \overline V)$.
     The incremental storage function (\ref{incrementalsf}) is defined as a Bregman distance   \cite{bregman_1967_CMMP} associated with (\ref{sf}) for the points $( \eta, f,  V)$ and $(\overline \eta, \overline f, \overline V)$. Due to the strict convexity of $S_1$ around $(\overline \eta, \overline f, \overline V)$, (\ref{incrementalsf}) has a local minimum at $(\overline \eta, \overline f, \overline V)$.
  \end{IEEEproof}
  \bigskip
%\smallskip
\begin{remark}({\bf Different power network models})
  The focus of this work is to achieve OLFC by distributed sliding mode control for the nonlinear power network, explicitly taking into account the turbine-governor dynamics. Equations (\ref{syscompact}) adequately represent a power network for the purpose of frequency regulation and are often further simplified by assuming constant voltages, leading to the so called `swing equations'. To the  analysis in this paper the incremental passivity property established above is essential, which has been derived for various other models, including microgrids. It is therefore expected that the presented approach can be straightforwardly applied to a wider range of models than the one we consider in this paper.   $\hfill \sq$
\end{remark}
\section{Optimal frequency regulation}\label{sec4}
In this section we formulate the control objectives of optimal load frequency control.
Before doing so, we first note that the steady state frequency $\overline f$, is generally different from zero without proper adjustments of $\overline u$ \cite{trip_2016_automatica}. %\smallskip
\bigskip
\begin{lemma}({\bf Steady state frequency})
  Let Assumption 2 hold, then necessarily $\overline f = \mathds{1}_n f^*$ with
\begin{align}\label{ssf}
  f^* = \frac{\mathds{1}_n^T (\overline u - P_{d} ) }{\mathds{1}_n^T(K_p^{-1} + R^{-1})\mathds{1}_n },
\end{align}
where $\mathds{1}_n \in \mathds{R}^n$ is the vector consisting of all ones.  $\hfill \sq$
\end{lemma}
%\smallskip
%\begin{IEEEproof}
%  The proof follows from algebraic manipulations of (\ref{sseq}) and (\ref{sstg}).
%\end{IEEEproof}
%\smallskip
%steady state frequency deviation $\overline f$ necessarily satisfies $\overline f = \mathds{1} f^*$ with
%\begin{align}
%  f^* = \frac{\mathds{1}^T (\overline u - P_{d} ) }{\mathds{1}^T(K_p^{-1} + R^{-1})\mathds{1} },
%\end{align}
\bigskip
 This leads us to the first objective,
concerning the regulation of the frequency deviation.
%\smallskip
\bigskip
\begin{objective}({\bf Frequency regulation})
  \begin{align}
  \lim_{t \rightarrow \infty} f(t)= \mathbf{0}.
\end{align}
$ \hfill \sq$
\end{objective}
%To formulate the second objective we differentiate between transmission lines $\mathcal{E}_s$ that have a \emph{scheduled} power flow and transmission lines $\mathcal{E}_f$ where the steady state power flow is \emph{flexible} (unscheduled). Note that $\mathcal{E} = \mathcal{E}_s \cup \mathcal{E}_f$ and that we can partition $\mathcal{B}$, without loss of generality, as
%\begin{align}\label{partition}
%  \mathcal{B} = \begin{bmatrix}
%    \mathcal{B}_s & \mathcal{B}_f
%  \end{bmatrix},
%\end{align}
%where  the matrices $\mathcal{B}_s \in \mathds{R}^{n \times e_s}$ and  $\mathcal{B}_f \in \mathds{R}^{n \times e_f}$ are obtained
%by collecting from $\mathcal{B}$ the columns indexed by $\mathcal{E}_s$ and $\mathcal{E}_f$ respectively.
%Control areas $\mathcal{V}_s \subseteq \mathcal{V}$ connected to an edge $k \in \mathcal{E}_s$ aim at maintaining the scheduled net power flows, leading to the following objective:
 %\smallskip
 \bigskip
From (\ref{ssf}) it is clear that it is sufficient that $\mathds{1}_n^T (\overline u - P_{d} ) =0 $, to have zero frequency deviation at the steady state. Therefore, there is flexibility to distribute the total required generation optimally among the various control areas. To make the notion of optimality explicit we assign to every control area a strictly convex linear-quadratic cost function $C_i(P_{ti})$ related to the generated power $P_{ti}$:
% Control areas $\mathcal{V}_f \subseteq \mathcal{V}$ interconnected by lines $k \in \mathcal{E}_f$ can freely adjust the steady state power flows between (some of) the areas and permit to optimize the associated generation costs required to obtain frequency regulation. To this end, we assign to every area a strictly convex linear-quadratic cost function
 \begin{align}\label{costfunction}
   C_i(P_{ti}) = ~ \frac{1}{2} \mathcal{Q}_i P_{ti}^2 + \mathcal{R}_i P_{ti} + \mathcal{C}_i \quad \forall i \in \mathcal{V}.
 \end{align}
% Note that, depending on the lines $\mathcal{E}_f$, the set $\mathcal{V}_f$ can be decomposed in $c$ distinct sets (clusters) $\mathcal{V}_f = \mathcal{V}_{f1} \cup \hdots \cup \mathcal{V}_{fc}$ that are not connected by lines in $\mathcal{E}_f$. Only within a cluster $V_{fi}$ the generation can be coordinated optimally. Again, without loss of generality, (\ref{partition}) can be partitioned such that
%\begin{align}
%\mathcal{B}_f = \begin{bmatrix}
%  \mathcal{B}_{f1} \\
%  \vdots \\
%  \mathcal{B}_{fc} \\
%  \boldsymbol{0}
%\end{bmatrix},
%\end{align}
% where  the matrices $\mathcal{B}_{fi} \in \mathds{R}^{n_{fi} \times e_f}$ are obtained
%by collecting from $\mathcal{B}_f$ the rows indexed by $\mathcal{V}_{fi}$.
%%\smallskip
%\begin{customobj}{3}({\bf Maintaining scheduled net power flows})
%  \begin{align}
%  \lim_{t \rightarrow \infty} \sum_{k \in \mathcal{N}_i} \mathcal{B}_{ik} \Gamma_k \sin(\eta_k(t))= \overline P_i^{net} \quad \forall i \in \mathcal{V}_a
%\end{align}
%\end{customobj}
%where $\overline P_i^{net}$ is the desired net power flow at control area $i \in \mathcal{V}_a$.
%In case the power network does not contain cycles, Objective 2 is equivalent to $\lim_{t \rightarrow \infty} \Gamma \sin(\eta(t))= \overline P_f$, such that the power flow on every line is identical to its desired value (see Remark \ref{remarkcycle} in Section 5). To be able to satisfy objectives 1 and 2, we make the following assumption on the feasibility of the control problem.
Minimizing the total generation cost, subject to the constraint that allows for a zero frequency deviation can then be formulated as the following optimization problem:
\begin{align}
\begin{split}\label{optimal}
  &\min \sum_{i \in \mathcal{V}} C_i(P_{ti}) \\
  \text{s.t.} \quad &  \mathds{1}_n^T (\overline u - P_{d} ) =0.
  \end{split}
\end{align}
The lemma below makes the solution to (\ref{optimal}) explicit \cite{trip_2016_automatica}:
%\smallskip
\bigskip
\begin{lemma}({\bf Optimal generation})
The solution $\overline P^{opt}_t$ to (\ref{optimal}) satisfies
\begin{align}\label{optimal.u}
\overline P^{opt}_t =~ \mathcal{Q}^{-1}(\overline \lambda^{opt} - \mathcal{R}),
\end{align}
where
\begin{align} \label{optimal.u2}
\overline \lambda^{opt} =~ \frac{\mathds{1}_n\mathds{1}_n^T (P_{d}+ \mathcal{Q}^{-1}\mathcal{R})}{\mathds{1}_n^T \mathcal{Q}^{-1}\mathds{1}_n}, %\in \mathcal{R}(\mathds{1}_{n_g}).
\end{align}
and $\mathcal{Q} = \text{diag}(\mathcal{Q}_1,\hdots, \mathcal{Q}_n)$, $\mathcal{R} = (\mathcal{R}_1,\hdots,\mathcal{R}_n)^T$.   $\hfill \sq$
\end{lemma}
\bigskip
% \medskip
From (\ref{optimal.u}) it follows that the marginal costs $\mathcal{Q}\overline P_t^{opt }+\mathcal{R}$ are identical. Note that (\ref{optimal.u}) depends explicitly on the \emph{unknown} power demand $P_{d}$. We aim at the design of a controller solving (\ref{optimal}) without measurements of the power demand, leading to the second objective.
%\smallskip
\bigskip
\begin{customobj}{2}({\bf Economic dispatch})
  \begin{align}
  \lim_{t \rightarrow \infty} P_t(t)= \overline  P_t^{opt},
\end{align}
with $\overline  P_t^{opt}$ as  in (\ref{optimal.u}), without measurements of $P_{d}$.   $\hfill \sq$
\end{customobj}
% \medskip
\bigskip
In order to achieve Objective 1 and Objective 2 we refine Assumption 2 that ensures the feasibility of the objectives.
%\smallskip
\bigskip
\begin{assumption}({\bf Existence of a optimal steady state}) Assumption 2 holds when $\overline f = \boldsymbol{0}$ and $\overline P_t = \overline P_g = \overline P_t^{opt}$, with $\overline P_t^{opt}$ as in (\ref{optimal.u}).   $\hfill \sq$
\end{assumption}
%\smallskip
\bigskip
\begin{remark}({\bf{Varying power demand}})
  To allow for a steady state solution, the power demand (unmatched disturbance) is required to be constant. This is not needed to reach the desired sliding manifold introduced in the next section, but is required only to establish the asymptotic convergence properties in Objective 1 and Objective 2. Furthermore, the proposed solution shows (\cite[Remark 8]{trip_2016_automatica}) the existence of a finite $\mathcal{L}_2$-to-$\mathcal{L}_\infty$ gain and a finite $\mathcal{L}_2$-to-$\mathcal{L}_2$ gain from a varying demand to the frequency deviation $\omega$ \cite{kundur2004}, once the system evolves on the sliding manifold, introduced in the next section.   $\hfill \sq$
\end{remark}
%%\smallskip
%\begin{customobj}{3}({\bf Minimizing generation costs})
%  \begin{align}
%  \lim_{t \rightarrow \infty} \mathcal{B}\Gamma \sin(\eta(t))= \mathcal{B} \overline P_f,
%\end{align}
%\end{customobj}
%\begin{remark}({\bf{Two particular cases}})
%  We point out two important particular cases of the presented setting that are commonly considered in the literature.
%  In the first case, every control area aims at maintaining its desired net power flow, i.e. $\mathcal{V}_s = \mathcal{V}$ and $\mathcal{V}_f = \emptyset$. In the second case, all areas participate in the \emph{optimal} LFC, i.e. $\mathcal{V}_s = \emptyset$ and $\mathcal{V}_f = \mathcal{V}$.
%\end{remark}
%Furthermore, we desire the controllers to be distributed and able to provide a continuous control input. We are now in a position to formulate the control problem for which we provide a solution in the next section.
%medskip
%\begin{problem}({\bf{Distributed control}})
%Let assumptions \ref{ass:0} and 2 hold. Given system \eqref{syscompact}, design a distributed control scheme, providing a continuous control input, capable of guaranteeing that the controlled system is asymptotically stable with zero steady state frequency deviation (Objective 1), maintaining, at the steady state, the scheduled (net) power flows (Objective 1) and minimizing the generation costs (Objective 3).
%\end{problem}
\section{Distributed sliding mode control}
\label{sec5}
In Section \ref{sec3} we discussed a passivity property of the power network (\ref{syscompact}), with input $P_t$ and output $f$. Unfortunately, the turbine-governor system (\ref{tgcompact}) does not immediately allow for a passive interconnection, since (\ref{tgcompact})  is a linear system with relative degree two, when considering $-f$ as the input and $P_t$ as the output\footnote{~A linear system with relative degree two is not passive, as follows e.g. from the Kalman-Yakubovich-Popov (KYP) lemma.}.
To alleviate this issue we propose  a \emph{distributed} Suboptimal Second Order Sliding Mode (D--SSOSM) control algorithm that simultaneously achieves Objective 1 and Objective 2, by constraining (\ref{tgcompact}) such that it enjoys a suitable passivity property, and by exchanging information on the marginal costs.
As a first step (see also Remark \ref{rem:firstorder} below), we augment the turbine-governor dynamics (\ref{tgcompact}) with a distributed control scheme, resulting in:
\begin{align}\label{eq:theta}
\begin{split}
T_t\dot{P}_{t}  = &-P_{t}  +  P_{g} \\
T_g\dot{P}_{g}  = &-R^{-1} f  - P_{g}   +  u \\
T_\theta \dot \theta =& -\theta + P_t - A\mathcal{L}^{com}(\mathcal{Q}\theta + \mathcal{R}).
\end{split}
\end{align}
Here, $\mathcal{Q}\theta + \mathcal{R}$ reflects the `virtual' marginal costs and $\mathcal{L}^{com}$ is the Laplacian matrix corresponding to the topology of an underlying communication network. The diagonal matrix $T_{\theta} \in \mathds{R}^{n \times n}$ provides
additional design freedom to shape the transient response and the matrix $A$ is suggested later to obtain a suitable passivity property.
We note that $\mathcal{L}^{com}(\mathcal{Q}\theta + \mathcal{R})$ represents the exchange information on the marginal costs among the control areas. To guarantee an optimal coordination of generation among \emph{all} the control areas the following assumption is made:
%\smallskip
\bigskip
\begin{assumption}({\bf{Communication topology}})\label{assum:com}
The graph corresponding to the communication topology is undirected and connected.
$ \hfill \sq$
%\footnote{~A graph is balanced if for every node $i \in \mathcal{V}$ its in-degree is equal to its out-degree. Every undirected graph is balanced.}
\end{assumption}
\bigskip
%\smallskip
%The topology of the communication network is allowed to be different than the topology of the power network, as long as Assumption \ref{assum:com} holds.
%\smallskip
\begin{remark}({\bf{First order turbine-governor dynamics}})\label{rem:firstorder}
  The rational behind this seemingly ad-hoc choice of the augmented dynamics is that for the controlled first order turbine-governor dynamics, where $u = \theta$ and $P_g = -R^{-1} f + \theta$, system
  \begin{align}\label{eq:theta2}
\begin{split}
T_t\dot{P}_{t}  = &-P_{t} -R^{-1} f   +  \theta\\
T_\theta \dot \theta =& -\theta + P_t - R^{-1}\mathcal{Q}\mathcal{L}^{com}(\mathcal{Q}\theta + \mathcal{R}),\\
\end{split}
\end{align}
  has been shown to be incrementally passive with input $-f$ and output $P_t$, and is able to solve Objective 1 and Objective~2~\cite{trip_2017_tns}. We aim at the design of $u$ and $A$ in (\ref{eq:theta}), such that (\ref{eq:theta}) behaves similarly as (\ref{eq:theta2}). This is made explicit in Lemma 5 and Lemma \ref{lemma4}. $ \hfill \sq$
  \end{remark}
% \medskip
\bigskip
To facilitate the discussion, we recall some definitions that are essential to sliding mode control.
To this end, consider system \begin{align} \begin{split}
  \dot x =&~ \zeta(x,u) \label{eq:smsystem}
  \end{split}
  \end{align}
  with $x \in \mathds{R}^n$, $u \in \mathds{R}^m$.
%\smallskip
\bigskip
\begin{definition}({\bf{Sliding function}})
\label{def:sliding_function}
The sliding function $\sigma(x): \R^{n} \rightarrow \R^m$ is a sufficiently smooth output function of system~\eqref{eq:smsystem}. $ \hfill \sq$
\end{definition}
%\smallskip
\bigskip
\begin{definition}{(\bf{$r$--sliding manifold})}
\label{def:sliding_manifold}
The $r$--sliding manifold\footnote{For the sake of simplicity, the order $r$ of the sliding manifold is omitted in the following.} is given by
%\begin{align}
%\label{eq:manifold_definition}
%  \{x \in \mathds{R}^n : \sigma(x) =  \sigma(x)^{(1)} = \dots = \sigma(x)^{(r-1)} = \boldsymbol{0}\},
%\end{align}
\begin{align}
\label{eq:manifold_definition}
  \{x \in \mathds{R}^n, u \in \mathds{R}^m : \sigma =  L_\zeta \sigma = \dots =L_\zeta^{(r-1)} \sigma = \boldsymbol{0}\},
\end{align}
where $L_\zeta^{(r-1)} \sigma(x)$ is the $(r-1)$-th order Lie derivative of $\sigma(x)$ along the vector field $\zeta(x,u)$. With a slight abuse of notation we also write  $L_\zeta \sigma(x) = \dot \sigma (x)$. $ \hfill \sq$
%where $\sigma(x)^{(r-1)}$ is the $(r-1)$-th order derivative of the sliding function $\sigma(x)$.
%The order of the sliding mode is defined as the largest value $r$, such that \eqref{eq:manifold_definition} holds for all $t \geq T_r$, $T_r$ being the time instant when the sliding manifold is reached. The order of a sliding mode controller is identical to the order of the sliding mode that it enforces.
\end{definition}
\bigskip
%\smallskip
\begin{definition}{(\bf{$r$--sliding mode})}
\label{def:sliding_mode}
A $r$--order sliding mode is enforced from $t = T_r \geq 0$, when, starting from an initial condition $x(0) = x_0$, the state of \eqref{eq:smsystem} reaches the $r$--sliding manifold \eqref{eq:manifold_definition}, and remains there for all $t\geq T_r$. $ \hfill \sq$
\end{definition}
% \medskip
\bigskip
Furthermore, the order of a sliding mode controller is identical to the order of the sliding mode that it is aimed at enforcing.
%\smallskip\\
We now propose a sliding function $\sigma(f,P_t,P_g,\theta)$ and a matrix $A$ for system (\ref{eq:theta}), which will allow us to prove convergence to the desired state. The choices are motivated by the stability analysis in the next section, but are stated here for the sake of exposition. First, the sliding function $\sigma: \R^{4n} \rightarrow \R^{n}$ is given by
\begin{align}
\begin{split}
\label{eq:sigma}
\sigma(f,P_t,P_g,\theta) = & ~M_1 f + M_2P_t + M_3P_g + M_4 \theta,
\end{split}
\end{align}
where $M_1 > \boldsymbol{0}$, $M_2 \geq \boldsymbol{0}$, $M_3 > \boldsymbol{0}$  are diagonal matrices and
$M_4 = -(M_2 + M_3)$. Therefore, $\sigma_i, i\in \mathcal{V}$, depends only on the locally available variables that are defined on node $i$, facilitating the design of a distributed controller (see Remark \ref{rem:distri}).
Second, the diagonal matrix $A \in \mathds{R}^{n \times n}$ is defined as
\begin{align}\label{eq:A}
    A = (M_2 + M_3)^{-1}M_1\mathcal{Q}.
\end{align}
By regarding the sliding function~\eqref{eq:sigma} as the output function of system \eqref{syscompact}, \eqref{eq:theta}, it appears that the relative degree
%\footnote{~The relative degree is the minimum order $\rho$ of the time derivative $\sigma_i^{(\rho)}, i \in \mathcal{V}$, of the sliding function associated to the $i$-th node in which the control $u_i,i\in \mathcal{V}$, explicitly appears.}
of the system is one.
This implies that a first order sliding mode controller can be {\textit{naturally}} applied \cite{Utkin} in order to attain in a finite time, the sliding manifold defined by $\sigma =\boldsymbol{0}$.
However, the input $u$ to the governor affects the first time derivative of the sliding function, i.e. $u$ affects $\dot \sigma$. Since sliding mode controllers generate a discontinuous signal, we additionally require $\dot \sigma = \boldsymbol{0}$, to guarantee that  the signal $u$ is continuous.
%Then, in order to provide a continuous control input $u$ to the governor, we also
Therefore, we define the desired sliding manifold as
 \begin{align}\label{manifold}
  \{(\eta,f,V,P_t,P_g,\theta): \sigma= \dot \sigma =\boldsymbol{0} \}.
\end{align}
%\todo{Can we say already something why the derivative appears and why we need it later on.?}
%It is by this choice of the sliding variable that the dynamics for the controller state $\theta$ are coupled to the power network.
%We now propose a specific sliding manifold, which indeed results in a suitable passivity property of (\ref{eq:theta}) as shown in Lemma 6 in the next section.
%%\smallskip
%\begin{assumption}\label{assum:sliding}({\bf{Desired sliding manifold}})
%  Let $M_1 > \boldsymbol{0}$, $M_2 \geq \boldsymbol{0}$, $M_3 > \boldsymbol{0}$  diagonal matrices and define
%$
%   M_4 = -(M_2 + M_3).
% $
%  I.e. the desired manifold is given by
%\begin{align} \label{manifold}
%  \{(\eta,\omega,P_t,P_g,\theta): M_1 f + M_2(P_t - \theta) + M_3(P_g - \theta) = \boldsymbol{0} \}.
%\end{align}
%\end{assumption}
%%The permitted values for $M_1, \hdots M_5$ follow from the stability analysis in
%
%%\todo{Add remark about other second order sliding mode controllers. PAssivity concept, idea holds for general class of sm controllers}
%%\todo{Add remark about the use of first order sm controllers}
%%\todo{Comment (more?) about adaptive gains}
%
%\begin{assumption}({\bf Choice of $A$})
%   Let
%$
%    A = (M_2 + M_3)^{-1}M_1\mathcal{Q}.
%$
%\end{assumption}
We continue by discussing a possible controller attaining the desired sliding manifold (\ref{manifold}) while providing a continuous control input $u$.
\subsection{Suboptimal Second Order Sliding Mode controller}
%\todo{adjust below. Maybe we can call $\rho$ the relative degree w.r.t. $u$ and $\tilde \rho$ the relative degree w.r.t. $\tilde u$ (now $w$).}

\begin{figure}[t]

        \centering
        \begin{tikzpicture}[scale=0.8, transform shape]

\node [text width=2.6cm, text height=4cm, fill=white, dashed, draw=gray, rectangle, rounded corners, text centered] at (0.7,0.25){ };
\draw (0.7,2.7) node
{Local information};

\node [text width=1.6cm, text height=2.3cm, fill=white, dashed, draw=gray, rectangle, rounded corners, text centered] at (3.4,-0.6){ };
\draw (3.4,1.4) node {Shared};
\draw (3.4,1) node {information};

        \bXInput{input}

        \begin{small}
        \bXCompSum[5]{sum1}{input}{+}{+}{+}{}
        \end{small}

        \bXLink[]{input}{sum1}
        \bXLinkName[1]{input}{$M_{2_{ii}}P_{ti}$}

        \bXBranchy[-4]{input}{d}
        \bXLinkxy[]{d}{sum1}
        \bXLinkName[1]{d}{$M_{1_{ii}}f_{i}$}

        \bXBranchy[4]{input}{d2}
        \bXLinkxy{d2}{sum1}
        \bXLinkName[1]{d2}{$M_{3_{ii}}P_{gi}$}

	\begin{small}
        \bXCompSum[7]{sum2}{sum1}{}{+}{+}{}
        \end{small}

        \bXLink[]{sum1}{sum2}

	\bXBranchy[4]{sum2}{d3}
        \bXLink[$M_{4_{ii}}\theta_{i}$]{d3}{sum2}
        %\bXLinkName[1]{d3}{$M_{4_{ii}}\theta_{i}$}

        \bXStyleBloc{rounded corners,fill=blue!10,text=black}
        \bXBloc[3.5]{controller}{SSOSM}{sum2}
        \bXLinkName[3.0]{controller}{}

        \bXLink[$\sigma_i$]{sum2}{controller}

        \bXBloc[2.5]{int}{\Large $\frac{1}{s}$}{controller}

        \bXLink[$w_i$]{controller}{int}

        \bXOutput[2.5]{output}{int}
        \bXLink[$u_i$]{int}{output}

        \end{tikzpicture}
        \caption{Block diagram of the proposed Distributed Suboptimal Second Order Sliding Mode (D--SSOSM) control strategy.}
        \label{fig:control_scheme}
\end{figure}
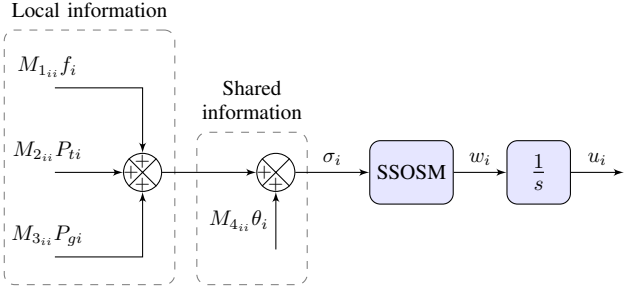

To prevent chattering, it is important to provide a continuous control input $u$ to the governor. Since sliding mode controllers generate a discontinuous control signal, we adopt the procedure suggested in \cite{661074} and first integrate the discontinuous signal, yielding for system (\ref{eq:theta}):
\begin{align}\label{eq:theta3}
\begin{split}
T_t\dot{P}_{t}  = &-P_{t}  +  P_{g} \\
T_g\dot{P}_{g}  = &-R^{-1} f  - P_{g}   +  u \\
T_\theta \dot \theta =& -\theta + P_t - A\mathcal{L}^{com}(\mathcal{Q}\theta + \mathcal{R}) \\
\dot u =&~ w,
\end{split}
\end{align}
where $w$ is the new (discontinuous) input generated by a sliding mode controller discussed below. A consequence is that the system relative degree (with respect to the new control input $w$) is now two, and we need to rely on a second order sliding mode control strategy to attain the sliding manifold (\ref{eq:sigma}) in a finite time \cite{Levant03}.
To make the controller design explicit, we discuss a specific second order sliding mode controller, the so-called `Suboptimal Second Order Sliding Mode' (SSOSM) controller proposed in \cite{661074}.
%We focus on the so-called `Suboptimal Second Order Sliding Mode' (SSOSM) controller \cite{661074}, since it furthermore permits to generate a continuous control input $u$ to the governor, as we will show below.
%\todo{ST: for me its not clear if the SM controller depends on $\sigma$ or on $\xi_1$. \color{blue}{$\sigma = \xi_1$}}
%\todo{ST: I messed up what the control and what the auxiliary system is. Basically, I don;t understand it. Can you fix it?}
%\todo{I removed $\sigma = \dot \sigma = 0$, as I don't understand where we use it except in the stability analysis. PRobably needs to be reintroduced. \color{blue}{Maybe we don't use it, but it is the definition of \lq\lq generation of SOSM''}}
%\todo{Can I understand from here that we need a second order sliding mode controller? \color{blue}{No you can't. You can from the relative degree (I will move it here)}}
  %scheme proposed in \cite{trip_2017_tns}, leading to an overall distributed solution.
%%\smallskip \\
%We now continue with describing the SSOM controller, based on  \cite{661074}, that guarantees finite time convergence to the sliding manifold $\sigma = \dot \sigma = {\bf{0}}$. To obtain a continuous control input (see Remark \ref{rem:input}),
We introduce two auxiliary variables $\xi_1=\sigma \in \mathds{R}^n$ and $\xi_2=\dot{\sigma} \in \mathds{R}^n$, and define the so-called auxiliary system as:
%and build the so-called auxiliary system as follows
\begin{eqnarray}
\label{eq:aux}
\begin{cases}
\dot{\xi}_1 = \xi_2\\
\dot{\xi}_2 = \phi(\eta,f,V,P_t,P_g,\theta) + G w.\\
\end{cases}
\end{eqnarray}
Bearing in mind  hat $\dot{\xi}_2 = \ddot \sigma = \phi + G w $, the expressions for the mapping $\phi$ and matrix $G$ can be straightforwardly obtained from (\ref{eq:sigma}) by taking the second derivative of $\sigma$ with respect to time, yielding for the latter\footnote{The expression for $\phi$ is rather long and is omitted.} $G = M_3T_g^{-1} \in \mathds{R}^{n \times n}$. We assume that the entries of $\phi$ and $G$ have known bounds
%\todo{The discussion on boundedness of the various states is not convincing. To be improved.}
\begin{equation}
\label{eq:boundF}
\abs{\phi_i} \leq \Phi_i \qquad \qquad \qquad \qquad ~\forall i \in \mathcal{V}
\end{equation}
\begin{equation}
\label{eq:boundG}
0<G_{\min_i} \leq G_{ii} \leq G_{\max_i}  \quad ~ \forall i \in \mathcal{V}
\end{equation}
with $\Phi_i, \, G_{\min_i}$ and $G_{\max_i}$ being positive constants. Second, $w$ is a discontinuous control input described by the SSOSM control algorithm \cite{661074}, and consequently for each area $i \in \mathcal{V}$, the control law $w_i$ is given by
\begin{equation}
\label{eq:SSOSM_law}
 w_i =-\alpha_i W_{\max_i} \sgn \left(\xi_{1_i}-\frac{1}{2}\xi_{1,\max_i}\right),
\end{equation}
with
\begin{equation}
\label{eq:constraint}
W_{\max_i}>\max\left(\frac{\Phi_i}{\alpha_i^{\ast}G_{\min_i}};\frac{4\Phi_i}{3G_{\min_i}-\alpha_i^{\ast}G_{\max_i}}\right),
\end{equation}
\begin{equation}
\label{eq:alpha}
\alpha_i^{\ast}\in (0, 1]\cap \biggl( 0, \frac{3 G_{\min_i}}{G_{\max_i}}\biggr),
\end{equation}
$\alpha_i$ switching between $\alpha_i^\ast$ and 1, according to \cite[Algorithm~1]{661074}. Note that indeed the input signal to the governor, $u(t) = \int_0^t w(\tau) d\tau$, is continuous, since the input $w$ is piecewise constant.
%\smallskip\\
The extremal values $\xi_{1,\max_i}$  in \eqref{eq:SSOSM_law} can be detected by implementing for instance a peak detection as in \cite{728381}.
{\color{black}{The block diagram of the proposed control strategy is depicted in Figure \ref{fig:control_scheme}.}}
\bigskip
 \begin{remark}({\bf{Uncertainty of $\phi$ and $G$}})
The mapping $\phi$ and matrix $G$ are uncertain due to the presence of the unmeasurable power demand $P_{d}$ and voltage angle $\theta$, and possible uncertainties in the system parameters.
In practical cases the bounds in \eqref{eq:boundF} and \eqref{eq:boundG} can be determined relying on data analysis and physical insights.
%However, in any practical operational region of the power network the states and demand remain bounded and in a vicinity of the sliding manifold the control input $u$ remains close to the so-called equivalent control \cite{728381}.
However, if these bounds cannot be a-priori estimated, the adaptive version of the SSOSM algorithm proposed in \cite{J_Incremona_16} can be used to dominate the effect of the uncertainties.$ \hfill \sq$
\end{remark}
% is applied to the augmented turbine-governor dynamics (\ref{eq:theta}).
\bigskip
\begin{remark}({\bf Distributed control})\label{rem:distri}
%Because $M_1, \dots, M_4$ in~\eqref{eq:sigma} are diagonal matrices, each sliding variable $\sigma_i$ is defined by only local variables at node~$i$.
Given $A$ in (\ref{eq:A}), the  dynamics of $\theta_i$ in \eqref{eq:theta} read for node $i \in \mathcal{V}$ as
  \begin{align}
  \begin{split} \nonumber
    T_{\theta i }\dot \theta_i =& -\theta_i + P_{ti}  \\
    &- \frac{\mathcal{Q}_iM_{1_{ii}}}{M_{2_{ii}}+M_{3_{ii}}} \sum_{j \in \mathcal{N}_j^{com}} (\mathcal{Q}_i \theta _i + \mathcal{R}_i - \mathcal{Q}_j \theta_j - \mathcal{R}_j),
    \end{split}
  \end{align}
  where $\mathcal{N}_j^{com}$ is the set of controllers connected to controller $i$. Furthermore, ($\ref{eq:SSOSM_law}$) depends only on $\sigma_i$, i.e. on states defined at node $i$. Consequently, the overall controller is indeed distributed and only information on marginal costs needs to be shared among connected controllers.$ \hfill \sq$
\end{remark}
%%\smallskip
%\begin{remark}({\bf{Continuous control input}})\label{rem:input}
%Note that $u(t) = \int_0^t w(\tau) d\tau$ is indeed continuous, which is essential to avoid chattering.
%\todo{Improve statement below}
%The use of a \emph{Second} Order Sliding Mode (SOSM) controller is essential to obtain a continuous control input $u$, avoiding chattering. Since the system relative degree\footnote{~The relative degree is the minimum order $r$ of the time derivative $\sigma_i^{(r)}, i \in \mathcal{V}$, of the sliding variable associated to the $i$-th node in which the control $u_i,i\in \mathcal{V}$ explicitly appears.} is equal to 1, then, in order to obtain a continuous control input, a SOSM control algorithm can be applied by artificially increasing the relative degree of the system.
%\end{remark}
%\smallskip
\bigskip
\begin{remark}({\bf Alternative SOSM controllers})
In this work we rely on the SOSM control law proposed in \cite{661074}. However, to constrain system \eqref{syscompact} augmented with dynamics \eqref{eq:theta3} on the sliding manifold (\ref{manifold}), where $\sigma = \dot{\sigma} = \bf{0}$, any other SOSM control law that does not need the measurement of $\dot{\sigma}$ can be used {\color{black}{(e.g. the super-twisting control \cite{doi:10.1080/00207179308923053})}}. An interesting continuation of the presented results is to study the performance of various SOSM controllers within the setting of (optimal) LFC.$ \hfill \sq$
\end{remark}
\bigskip
\begin{remark}({\bf Comparison with \cite{trip_2017_tns} and \cite{kasis2017stability}})\label{rem:contri}
  The controller proposed in \cite{trip_2017_tns} requires, besides a gain restriction in the controller, that
       \begin{align}\label{assumineq}
     \begin{split}
     4T_{gi}T_{ti}^{-1} &> 1\\
     K_{pi}^{-1}T_{gi}T_{ti}^{-1} &> 1.
     \end{split}
     \end{align}
     In this work, we do not impose such restriction on the parameters. The result in \cite{kasis2017stability} requires, besides some assumptions on the dissipation inequality related to the generation side, the existence of frequency dependent generation and load, where the generation/demand (output) depends directly (e.g. proportionally) on the frequency (input), avoiding complications arising from generation dynamics that have relative degree two when considering the input-output pair just indicated (see also Remark \ref{rem:reldegree}).
\end{remark}
\bigskip
\begin{remark}({\bf{Primal-dual based approaches}})\label{rem:pd1}
 Although the focus in this work is to augment the power network with consensus-type dynamics in (\ref{eq:theta}), it is equally possible to augment the power network with a continuous primal-dual algorithm that has been studied extensively to obtain optimal LFC. This work provides therefore also means to extend existing results on primal-dual based approaches to incorporate the turbine-governor dynamics, generating the control input by a higher order sliding mode controller. The required adjustments follow similar steps as discussed in \cite[Remark 9]{trip_2017_tns}, and, for the sake of brevity, we directly state the resulting primal-dual based augmented system, replacing (\ref{eq:theta}),
 \begin{align}\label{eq:thetapd}
\begin{split}
T_t\dot{P}_{t}  = &-P_{t}  +  P_{g} \\
T_g\dot{P}_{g}  = &-R^{-1} f  - P_{g}   +  u \\
T_\theta \dot \theta =& -\theta + P_t - M_1(M_2 + M_3)^{-1}\bigg(\nabla C(\theta) - \lambda \bigg) \\
\dot v =&-\mathcal{B}^T\lambda \\
  \dot \lambda =& ~ \mathcal{B}v - \theta + P_d.
\end{split}
\end{align}

  In this case only strict convexity  of $C(\cdot)$ is required and the load $P_d$ explicitly appears in (\ref{eq:thetapd}). The stability analysis of the power network, including the augmented turbine-governor dynamics (\ref{eq:thetapd}), follows \emph{mutatis mutandis}, the same argumentation as in the next section where the focus is on the augmented system (\ref{eq:theta}). Some required nontrivial modifications in the analysis are briefly discussed in Remark \ref{rem:pd2}. $ \hfill \sq$
\end{remark}

\section{Stability analysis and main result} \label{sec6}
In this section we study the stability of the proposed control scheme, based on an enforced passivity property of (\ref{eq:theta}) on the sliding manifold defined by \eqref{eq:sigma}. First, we establish that the second order sliding mode controller \eqref{eq:aux}--\eqref{eq:alpha} constrains the system in finite time to the desired sliding manifold.
%%\smallskip
%\begin{assumption}\label{assum:sliding}({\bf{Desired sliding manifold}})
%  Let $M_1 > 0$, $M_2 \geq 0$, $M_3 >0$  diagonal matrices and let $M_4$ be defined as
%  \begin{align}
%  \begin{split}
%   M_4 =& -(M_2 + M_3), \\
%    \end{split}
%  \end{align}
%   in (\ref{eq:sigma}).
%  Furthermore, let
%  \begin{align}
%    A = (M_2 + M_3)^{-1}M_1Q,
%  \end{align}
%  (\ref{eq:theta}).
%\end{assumption}
%\smallskip
\bigskip
\begin{lemma}({\bf{Convergence to the sliding manifold}})\label{lemma3}
Let Assumption 1 hold. The solutions to system (\ref{syscompact}), augmented with \eqref{eq:theta3}, in closed loop with controller \eqref{eq:aux}--\eqref{eq:alpha}  converge in a finite time $T_{r}$ to the sliding manifold (\ref{manifold}) such that
  \begin{align}
\label{eq:Pg}
\begin{split}
P_g =& - M_3^{-1} (M_1f + M_2 P_t + M_4\theta) \quad \forall t \geq T_r.
 \end{split}
\end{align}
\end{lemma}
\begin{IEEEproof}
 %$\sigma(t) = \dot{\sigma}(t) = {\bf{0}}, \ \forall \, t \geq T_{r}$,
 Following \cite{661074}, the application of \eqref{eq:aux}--\eqref{eq:alpha}  to each control area guarantees that $\sigma = \dot{\sigma} = {\bf{0}}, \ \forall \, t \geq T_{r}$. The details are omitted, and are an immediate consequence of the used SSOSM control algorithm \cite{661074}.
   %a second order sliding mode is enforced, i.e. $\exists \; t_{r} \geq t_0 \, :  \ \sigma(t) = \dot{\sigma}(t) = {\bf{0}}, \ \forall \, t \geq t_{r}$, where $t_0$ and $t_{r}$ are the initial time instant and the reaching time, respectively.
Then, from (\ref{eq:sigma}) one can easily obtain \eqref{eq:Pg}, where $M_3$ is indeed invertible.
\end{IEEEproof}
%\smallskip
%The restrictions on $M_i$ and $\overline \eta$ are required to apply LaSalle's invariance principle in Theorem 1, where stability of the proposed control scheme is proven. This shows how the sliding manifold can be designed relying on an energy (storage) function based stability analysis. Before discussing this main result, some useful intermediate results are derived.
%First, we show that the second order sliding mode controller \eqref{eq:sigma}--\eqref{eq:alpha} constrains the system in finite time to the manifold characterized in the lemma below.
%\
\bigskip
Exploiting relation (\ref{eq:Pg}), on the sliding manifold where $\sigma = \dot \sigma= {\bf{0}}$, the so-called equivalent system is as follows:
\begin{align}\label{eq:equiv}
\begin{split}
M_3T_t\dot{P}_t  =& -(M_2+M_3)P_t - M_4 \theta - M_1 f \\
T_\theta \dot \theta =& -\theta + P_t  - A \mathcal{L}^{com}(\mathcal{Q}\theta + \mathcal{R}).\\
\end{split}
\end{align}
 As a consequence of the feasibility assumption (Assumption~\ref{a2}), the system above admits the following steady state:
\begin{align}\label{eqss}
  \begin{split}
    \boldsymbol{0} =& -(M_2+M_3)\overline P_t^{opt} - M_4 \overline \theta - M_1 \boldsymbol{0}  \\
 \boldsymbol{0}  =& -\overline \theta + \overline P_t^{opt}  - A \mathcal{L}^{com}(\mathcal{Q} \overline \theta + \mathcal{R}).\\
  \end{split}
\end{align}
Now, we show that system (\ref{eq:equiv}), with $A$ as in (\ref{eq:A}), indeed possesses a passivity property with respect to the steady state (\ref{eqss}).
Note that, due to the discontinuous control law \eqref{eq:SSOSM_law}, the solutions to the closed loop system are understood in the sense of Filippov. Following the equivalent control method \cite{Utkin}, the solutions to the equivalent system are however continuously differentiable.
\bigskip
\begin{lemma}({\bf Incremental passivity of (\ref{eq:equiv})}) \label{lemma4}
System (\ref{eq:equiv}) with input $-f$ and output $P_t$ is an incrementally passive system, with respect to the constant $(\overline P_t^{opt}, \overline \theta)$ satisfying (\ref{eqss}).
 \end{lemma}
  \vspace{0.2cm}
\begin{IEEEproof}
Consider the following incremental storage function
\begin{align}
\begin{split}
\label{eq:V2}
\mathcal{S}_2 = & ~\frac{1}{2}(P_t - \overline{P}^{opt}_t)^T  M_1^{-1}M_3 T_t (P_t - \overline{P}^{opt}_t)\\
& + \frac{1}{2}(\theta - \overline{\theta})^T  M_1^{-1}(M_2 + M_3)T_{\theta} (\theta - \overline{\theta}),
\end{split}
\end{align}
which is positive definite, since $M_1 >\boldsymbol{0}, M_2 \geq \boldsymbol{0}$ and $M_3 >\boldsymbol{0}$.
Then, we have that $\mathcal{S}_2$ satisfies along the solutions to (\ref{eq:equiv})
\begin{align}
\begin{split}
\label{eq:dV2} \nonumber
\dot{\mathcal{S}}_2 = &~ \frac{1}{2}(P_t- \overline{P}^{\, \mathrm{opt}}_{t})^T  M_1^{-1}M_3 T_{t} \dot P_{t} \\
& + \frac{1}{2}(\theta - \overline{\theta})^T  M_1^{-1}(M_2 + M_3)T_{\theta} \dot \theta \\
= & ~\frac{1}{2}(P_{t} - \overline{P}^{{opt}}_{t})^T (-M_1^{-1}(M_2+M_3)  P_{t}  - f - M_1^{-1}M_4 \theta)\\
& + \frac{1}{2}(\theta - \overline{\theta})^T  M_1^{-1}(M_2 + M_3)\\&\quad  \cdot(P_{t} -\theta - A\mathcal{L}^{com}(\mathcal{Q}\theta + \mathcal{R})).
\end{split}
\end{align}
In view of $M_4 = -(M_2 + M_3)$, $A = (M_2 + M_3)^{-1}M_1\mathcal{Q}$ and equality (\ref{eqss}), it follows that
\begin{align} \nonumber
  \begin{split}
    \dot{\mathcal{S}}_2 =& -(P_t - \theta)^TM_1^{-1}(M_2 + M_3)(P_t - \theta) \\
    &-(\mathcal{Q}\theta + \mathcal{R} - Q\overline \theta -R )\mathcal{L}^{com}(\mathcal{Q}\theta + \mathcal{R} - \mathcal{Q}\overline \theta - \mathcal{R}) \\
   & -(P_t - \overline P_t^{opt})^T(f - \boldsymbol{0}).
  \end{split}
\end{align}
%\begin{align}
%\begin{split}
%  \dot U_2 =&  -  (P_{\mathrm{t}} - \vartheta)^TM_1^{-1}(M_2 + M_3) (P_{\mathrm{t}} - \vartheta) \\
%  &-(Q\vartheta + Z)^T L_{\mathrm{c}}(Q\vartheta + Z) - (P_{\mathrm{t}} - \overline P_{\mathrm{t}}^{\, \mathrm{opt}})^Tf .
%  \end{split}
%\end{align}
\end{IEEEproof}
\bigskip
\begin{remark}({\bf Reducing the relative degree})\label{rem:reldegree}
  An important consequence of the proposed sliding mode controller \eqref{eq:aux}--\eqref{eq:alpha} is that the relative degree of system (\ref{eq:equiv}) is one with input $-f$ and output $P_t$. This is in contrast to the `original' system (\ref{tgcompact}) that has relative degree two with the same input--output pair.

  $ \hfill \sq$
\end{remark}
% \medskip
\bigskip
Now, relying on the interconnection of incrementally passive systems, we can prove the main result of this paper concerning the evolution of the augmented system controlled via the proposed distributed SSOSM control strategy.
%\smallskip
\bigskip
\begin{theorem}({\bf Main result: distributed OLFC})
\label{th:2}
Let assumptions \ref{ass:0}--6 hold.
Consider system \eqref{syscompact} and \eqref{eq:theta}, controlled via \eqref{eq:aux}--\eqref{eq:alpha}. Then, the solutions to the closed-loop system starting in a neighbourhood of the equilibrium $(  \overline{\eta}, \overline f = \boldsymbol{0}, \overline {V}, \overline{P}_t^{opt}, \overline{P_g}, \overline \theta)$ approach the set where $\overline{f}={\bf{0}}$ and  $\overline P_t = \overline P_t^{opt}$, with $\overline P_t^{opt}$ given by (\ref{optimal.u}).
\end{theorem}
 \vspace{0.2cm}
\begin{IEEEproof}
Following Lemma \ref{lemma3}, we have that the SSOSM control enforces system (\ref{eq:theta}) to evolve $\forall \, t \geq T_{r}$ on the sliding manifold (\ref{manifold}), resulting in the reduced order system \eqref{eq:equiv}.
To study the obtained closed loop system, consider the overall incremental storage function $\mathcal{S} = \mathcal{S}_1 + \mathcal{S}_2$, with $\mathcal{S}_1$ given by (\ref{incrementalsf}) and $\mathcal{S}_2$ given by (\ref{eq:V2}).
In view of Lemma 2, we have that $\mathcal{S}$ has a local minimum at $( \overline{\eta}, \overline{f}={\bf{0}}, \overline V, \overline{P}_t^{opt}, \overline{\theta})$ and satisfies along the solutions to \eqref{syscompact}, (\ref{eq:equiv})
%\todo{Check the term involving $\dot V$}
\begin{align}
\begin{split} \nonumber
    \dot{\mathcal{S}} =& -f^T K_p^{-1} f   - \dot V^T T_V(X_d - X'_d)^{-1} \dot V\\
    &-(P_t - \theta)^TM_1^{-1}(M_2 + M_3)(P_t - \theta) \\
    &-(\mathcal{Q}\theta + \mathcal{R} - Q\overline \theta -R )\mathcal{L}^{com}(\mathcal{Q}\theta + \mathcal{R} - \mathcal{Q}\overline \theta - \mathcal{R}) \\
     \leq &~ 0,
\end{split}
\end{align}
where $\dot V = T_V^{-1}\big(-(X_d - X'_d)E(\eta)V + \overline E_{f}\big)$.
Consequently, there exists a forward invariant set $\Upsilon$ around $( \overline{\eta}, \overline{f}={\bf{0}}, \overline V, \overline{P}_t^{opt}, \overline{\theta})$ and by LaSalle's invariance principle the solutions that start in $\Upsilon$ approach the largest invariant set contained in
\begin{align}
 & \Upsilon \cap \{(\eta, f, V, P_t, \theta): f = {\bf{0}}, V = \big( (X_d - X'_d)E(\overline \eta)\big)^{-1} \overline E_f, \nonumber\\& \hspace{8.9em}P_t = \theta, \theta = \overline \theta + \mathcal{Q}^{-1}\mathds{1}\alpha\},
\end{align}
where $\alpha \in \mathds{R}$ is some scalar.
On this invariant set the controlled power network satisfies
\begin{align}\label{sysinv}
\begin{split}
\dot{\eta}  = & ~\mathcal{B}^{T}\boldsymbol{0} \\
\boldsymbol{0}  = &  ~K_p( \overline \theta + \mathcal{Q}^{-1}\mathds{1}\alpha  - P_{d} - \mathcal{B}  \Gamma(V) \sin(  \eta ) )\\
\boldsymbol{0} =& -(X_d - X'_d)E(\eta)V + \overline E_{f} \\
M_3T_t\dot{P}_t  =& ~\boldsymbol{0} \\
T_\theta \dot \theta =&~ \boldsymbol{0}.\\
\end{split}
\end{align}
Pre-multiplying both sides of the second line of (\ref{sysinv}) with $\mathds{1}^T_nK_p^{-1}$ yields
$
  0 = \mathds{1}_n^T( \overline \theta + \mathcal{Q}^{-1}\mathds{1}\alpha  - P_{d}).
$
Since $\overline \theta = \overline P_t^{opt}$, $\mathds{1}_n^T (\overline P_t^{opt} - P_{d}) =0$ and $\mathcal{Q}$ is a diagonal matrix with only positive elements, it follows that necessarily $\alpha = 0$. We can conclude that the solutions to the system \eqref{syscompact} and \eqref{eq:theta}, controlled via \eqref{eq:aux}--\eqref{eq:alpha}, indeed approach the set where $\overline{f}={\bf{0}}$ and $\overline P_t = \overline P_t^{opt}$, with $\overline P_t^{opt}$ given by (\ref{optimal.u}).
%Bearing in mind that $\mathcal{B}\Gamma \sin(\overline \eta) = \mathcal{B}\overline P_f$, we can indeed observe
%hat system (\ref{syscompact})  approaches the set where the frequency deviation is zero, and where the net exchanged power is equal to the desired value, i.e. $ \mathcal{B}\Gamma \sin(\eta) = \mathcal{B}\overline P_f$.
\end{IEEEproof}
\bigskip
%%\smallskip
%\begin{remark}\label{remarkcycle}
%  ({\bf{Acyclic network topologies.}}) In case the topology of the power network does not contain any cycles, we have that the corresponding incidence matrix $\mathcal{B}$ has full column rank and therefore has a left-inverse satisfying $\mathcal{B}^{+}\mathcal{B} = I$, such that we can conclude from Theorem 1 that the system approaches the set where
%  \begin{align}
%  \begin{split}
%   \mathcal{B}\Gamma \sin(\overline \eta) =&\mathcal{B} \overline P_f\\
%    \mathcal{B}^+ \mathcal{B}\Gamma \sin(\overline \eta) =& \mathcal{B}^+\mathcal{B} \overline P_f.\\
%    \Gamma \sin(\overline \eta) =& \overline P_f.
%    \end{split}
%  \end{align}
%\end{remark}
%\smallskip
\begin{remark}({\bf{Robustness to failed communication}})
The proposed control scheme is distributed and as such requires a communication network to share information on the marginal costs.
 However, note that the term $- A\mathcal{L}^{com}(\mathcal{Q}\theta + \mathcal{R})$ in (\ref{eq:theta}) is not needed to enforce the  passivity property established in Lemma \ref{lemma4},
    but is required to prove convergence to the economic efficient generation $\overline{P}^{{opt}}_t$. In fact, setting $A = \boldsymbol{0}$ still permits to infer frequency regulation following the argumentation of Theorem 1. $ \hfill \sq$
\end{remark}
\bigskip
% \medskip
%\todo{Remark below to be adjusted after boundedness modifications}
%\begin{remark}({\bf{Tuning of control gains}})
% To achieve a desirable transient response of the power network the `gain' matrices $M_1, \dots, M_3$ in Assumption \ref{assum:sliding} can be tuned. The optimal tuning of the gain matrices is outside the scope of this paper and largely depends on the particular power network under consideration.
%\end{remark}
%%\smallskip
%  \todo{Antonella said that the Remark it's ok even if it is quite general. Anyway she believes that by properly shaping the manifold and by properly choosing the control gain, the Remark is reasonable.}
\begin{remark} ({\bf{Region of attraction}})
  LaSalle's invariance principle can be applied to all bounded solutions. As follows from Lemma 2, we have that the considered incremental storage function has a local minimum at the desired steady state, whereas the time to converge to the sliding manifold can be made arbitrarily small by properly choosing the gains of the SSOSM control. This guarantees that solutions starting in the vicinity of the steady state of interest remain bounded. A preliminary (numerical) assessment indicates that the region of attraction is large, but a thorough analysis is left as future endeavour. $ \hfill \sq$
   %A thorough analysis of the region of attraction is offers an interesting future direction, e.g by incorporating recent results in (\cite{vu_2016_tps}, \cite{dvijotham_2015_arxiv}) where energy functions, similar to the one used in this paper, are further characterized.
  %\todo{Integral SM control}
\end{remark}
\bigskip
\begin{remark} ({\bf{Stability of primal-dual based approaches}}) \label{rem:pd2}
To accommodate the additional dynamics of states $v$ and $\lambda $ appearing in primal-dual based augmented system (\ref{eq:thetapd}), an additional storage term is required in Lemma 6, namely:
  \begin{align}
    \mathcal{S}_3 =& ~\frac{1}{2}(v - \overline v)^T(v - \overline v)  +  \frac{1}{2}(\lambda - \overline \lambda)^T(\lambda - \overline \lambda),
  \end{align}
  where $\overline v$ and $\overline \lambda$ satisfy the steady state equations
   \begin{align}
\begin{split}
\boldsymbol{0} =& -\overline \theta + \overline P^{opt}_t - M_1(M_2 + M_3)^{-1}\bigg(\nabla C(\overline \theta) - \overline \lambda \bigg) \\
\boldsymbol{0} =&-\mathcal{B}^T \overline \lambda \\
\boldsymbol{0}  =& ~ \mathcal{B}\overline v - \overline \theta + P_d.
\end{split}
\end{align}
  Consequently, $\mathcal{S}_2 + \mathcal{S}_3$ satisfies along the solutions to the system, constrained to the manifold $\sigma = \dot \sigma = \boldsymbol{0}$,
  \begin{align} \nonumber
  \begin{split}
    \dot{\mathcal{S}}_2 + \dot{\mathcal{S}}_3=& -(P_t - \theta)^TM_1^{-1}(M_2 + M_3)(P_t - \theta) \\
    &-(\theta - \overline \theta)^T (\nabla C(\theta) - \nabla C(\overline \theta) )\\
   & -(P_t - \overline P_t^{opt})^T(f - \boldsymbol{0}).
  \end{split}
\end{align}
Note that, as a result of the mean value theorem, $-(\theta - \overline \theta)^T (\nabla C(\theta) - \nabla C(\overline \theta) ) = -(\theta - \overline \theta)^T \nabla^2C(\tilde \theta) (\theta - \overline \theta) \leq 0,$ for some $\tilde \theta_i \in [\theta_i, \overline \theta_i] $, for all $i \in \mathcal{V}$. The matrix $\nabla^2C(\tilde \theta) \in \mathds{R}^{n \times n}$ is positive definite due to the strict convexity of $C(\cdot)$.
  The proof of Theorem 1 can now be repeated using the incremental storage function $\mathcal{S} = \mathcal{S}_1 + \mathcal{S}_2 + \mathcal{S}_3$. $ \hfill \sq$
\end{remark}

\section{Case study}\label{sec7}
%\todo{Adjust, and can we have a power generation starting above zero?, i.e. not the incremental version}
\label{sec:simulations}
\begin{figure}
\begin{small}
\begin{center}
\begin{tikzpicture}[>=stealth',shorten >=1pt,auto,node distance=2.5cm,
                    semithick]
  \tikzstyle{every state}=[circle,thick,draw=black,fill=black!3,text=black]

  \node[state] (A)                    {Area 1};
  \node[state]         (B) [above right of=A] {Area 2};
  \node[state]         (D) [below right of=A] {Area 4};
  \node[state]         (C) [below right of=B] {Area 3};

  \path[->] (A) edge             node {$P_{12}$} (B)
  		edge	[left]	     node {$P_{14}$} (D)
           (B) edge              node {$P_{23}$} (C)
           (C) edge              node {$P_{34}$} (D);

  \path[<->] (A) edge [bend right, dashed, blue]          node {} (B)
  		%edge	 [bend right, dashed, red]	     	node {} (D)
           (B) edge [bend right, dashed, blue]          	node {} (C)
           (C) edge [bend right, dashed, blue]          	node {} (D);
\end{tikzpicture}
\caption{Scheme of the considered power network partitioned into 4 control areas, where $P_{ij}= \frac{V_{i} \, V_{j}} {X_{ij}} \sin{(  \delta_i  -   \delta_j )}$. The solid arrows indicate the positive direction of the power flows through the power network, while the dashed lines represent the communication network.}
\label{fig:microgrid_example}
\end{center}
\end{small}
\end{figure}
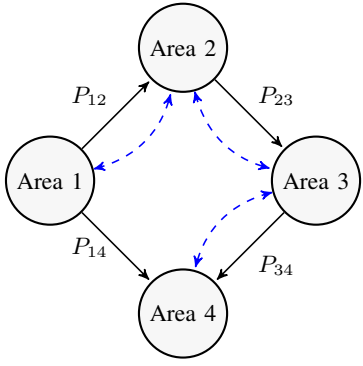
\vspace{0.1cm}
\begin{figure}
\begin{center}
\includegraphics[width=\columnwidth]{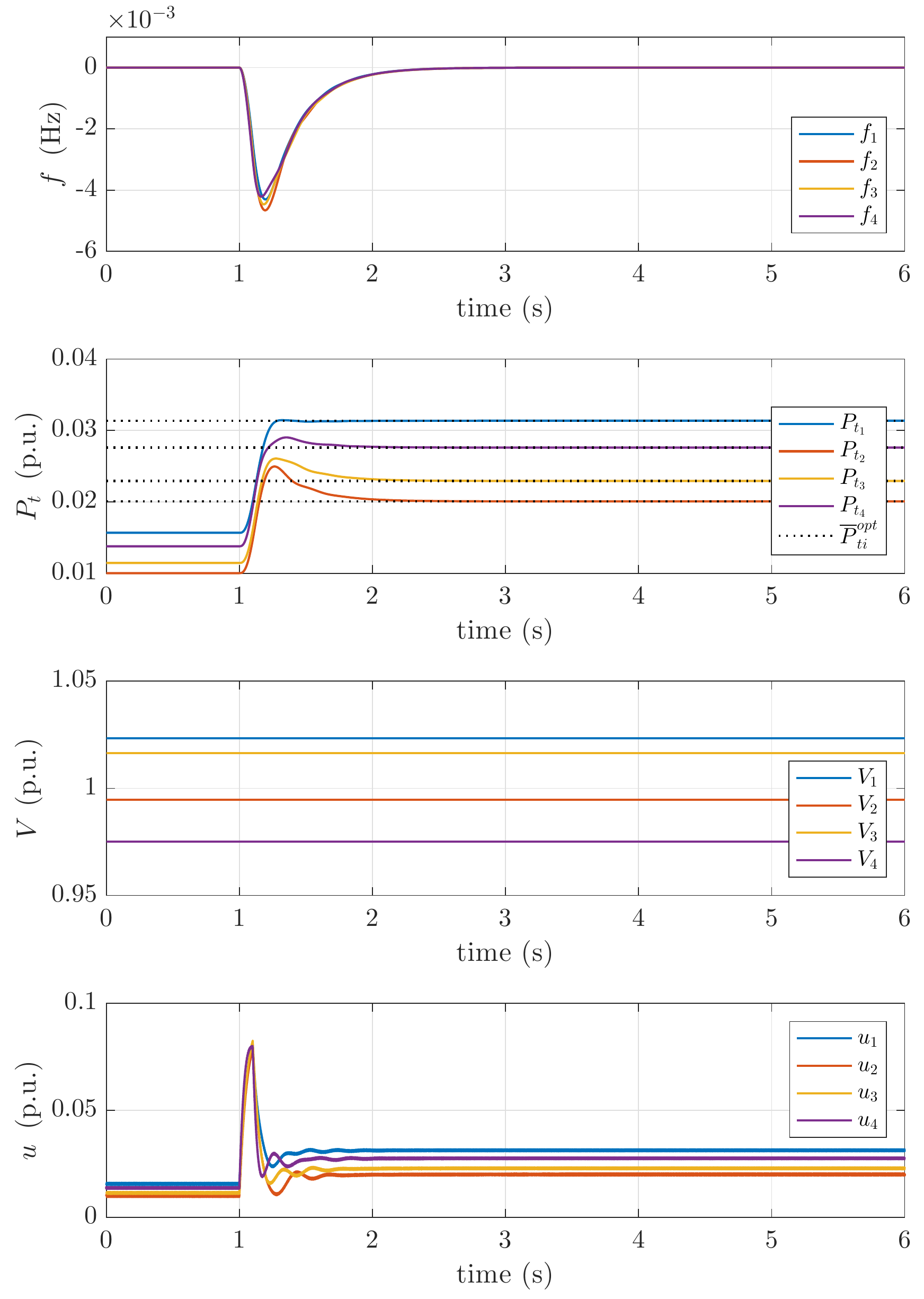}
\caption{Time evolution of the frequency deviation, generated power, voltage dynamics and control input considering a power demand variation at the time instant $t=$ \SI{1}{\second}.}
\label{fig:sf}
\end{center}
\end{figure}
In this section, the proposed control solution is assessed in simulation, by implementing a power network partitioned into four control areas (e.g. the IEEE New England 39-bus system \cite{6579989}). The topology of the power network is represented in Figure \ref{fig:microgrid_example}, together with the communication network (dashed lines).
\begin{table}
\caption{Network Parameters and Power Demand}
\centering
{\begin{tabular}{ll | cccc}			
								&	&\rotatebox{90}{Area 1}		&\rotatebox{90}{Area 2}		&\rotatebox{90}{Area 3}		&\rotatebox{90}{Area 4}\\					
\hline
 & & & & &\\
		$T_{pi}$	&(s)		&\num{21.0}		&\num{25.0}		&\num{23.0}		&\num{22.0}\\
		$T_{ti}$	&(s)		&\num{0.30}		&\num{0.33}		&\num{0.35}		&\num{0.28}\\
		$T_{gi}$	&(s)		&\num{0.080}		&\num{0.072}		&\num{0.070}		&\num{0.081}\\
		$T_{Vi}$	&(s)		&\num{5.54}		&\num{7.41}		&\num{6.11}		&\num{6.22}\\
		$K_{pi}$ 	&(Hz p.u.$^{-1}$)	&\num{120.0}		&\num{112.5}		&\num{115.0}		&\num{118.5}\\
		$R_i$	&(Hz p.u.$^{-1}$)	&\num{2.5}		&\num{2.7}		&\num{2.6}		&\num{2.8}\\
		$X_{di}$	&(p.u.)	&\num{1.85}		&\num{1.84}		&\num{1.86}		&\num{1.83}\\
		$X'_{di}$	&(p.u.)	&\num{0.25}		&\num{0.24}		&\num{0.26}		&\num{0.23}\\
		$\overline E_{fi}$	&(p.u.)	&\num{1.0}		&\num{1.0}		&\num{1.0}		&\num{1.0}\\
		$B_{ii}$	&(p.u.)	&\num{-13.6}		&\num{-12.9}		&\num{-12.3}		&\num{-12.3}\\
		$T_{\theta i}$	&(s)		&\num{0.33}		&\num{0.33}		&\num{0.33}		&\num{0.33}\\
		$\mathcal{Q}_i$	&(\num{e4} \$ h$^{-1}$)	&\num{2.42}		&\num{3.78}		&\num{3.31}		&\num{2.75}\\
$\Delta P_{di}$ &(p.u.) &\num{0.010} &\num{0.015} &\num{0.012} &\num{0.014}
\end{tabular}}
\label{tab:parameters}
\end{table}
The line parameters are $B_{12} = -5.4$ p.u., $B_{23} = -5.0$ p.u., $B_{34} = -4.5$ p.u. and $B_{14} = -5.2$ p.u., while the network parameters and the power demand $\Delta P_{di}$ of each area are provided in Table \ref{tab:parameters}, where a base power of \SI{1000}{M\watt} is assumed.
The matrices in \eqref{eq:sigma} are chosen as $M_1 = 3 I_4, \, M_2 = I_4, \, M_3 = 0.1 I_4$ and $M_4 = -(M_2+M_3)$, $I_4 \in \R^{4 \times 4}$ being the identity matrix, while the control amplitude $W_{\max_i}$ and the parameter $\alpha^{\ast}_i$, in \eqref{eq:SSOSM_law} are \num{10} and  \num{1}, respectively, for all $i \in \mathcal{V}$.
For the sake of simplicity, in the cost function \eqref{costfunction}, we select $\mathcal{R}_i=\mathcal{C}_i=0$ for all $i \in \mathcal{V}$.
The system is initially at the steady state.
Then, at the time instant $t=$ \SI{1}{\second}, the power demand in each area is increased according to the values reported in Table \ref{tab:parameters}.
From Figure \ref{fig:sf}, one can observe that the frequency deviations converge asymptotically to zero after a transient where the frequency drops because of the increasing load.
Indeed, one can note that the proposed controllers increase the power generation in order to reach again a zero steady state frequency deviation.
Moreover, the total power demand is shared among the areas, minimizing the total generation costs. More precisely, by applying the proposed D-SSOSM, the total generation costs are \num{10} \% less than the generation costs when each area would produce only for its own demand.
%
%%%%%%%%%% CONCLUSIONS
\section{Conclusions}
\label{sec:conclusions}
A Distributed Suboptimal Second Order Sliding Mode (D-SSOSM) control scheme is proposed to solve an optimal load frequency control problem in power systems. In this work, we adopted a nonlinear model of a power network, including voltage dynamics, where each control area is represented by an equivalent generator including second order turbine-governor dynamics. Based on a suitable chosen sliding manifold, the controlled turbine-governor system, constrained to this manifold, possesses an incremental passivity property that is exploited to prove that the frequency deviation asymptotically approaches zero and an economic dispatch is achieved.
%An important feature of the proposed distributed control approach is that the controllers do not require the measurement of the power demand nor rely on load observers.
Designing the sliding modes, based on passivity considerations, appears to be powerful and we will pursue this approach within different settings, such as achieving power sharing in microgrids. Additionally, we would like to compare the performance of the proposed sliding mode based control scheme with other approaches to OLFC appearing in the literature.
\balance
\bibliographystyle{IEEEtran}
\bibliography{LFC1}
%\appendix
%I removed the defintion of relative degree. It can be added, but then it needs to be earlier, as I also discuss the relative degree of a linear system.
\end{document}